\documentclass{article}
\usepackage{spconf,amsmath,graphicx}

\usepackage{enumitem}
\usepackage{amssymb}
\setlist{nosep, leftmargin=14pt}

\usepackage{mwe} 

\title{Hybridization of attention unet with repeated atrous spatial pyramid pooling for improved brain tumor segmentation}
\vspace{-2pt} 

\name{Satyaki Roy Chowdhury$^{1}$ and Golrokh Mirzaei$^{2}$}
\address{$^{1}$Department of Electrical and Computer Engineering, \\
$^{2}$Department of Computer Science and Engineering, \\
The Ohio State University, Columbus, USA}
\vspace{3pt} 

\begin{document}
%
\maketitle
\begin{abstract}
Brain tumors are highly heterogeneous in terms of their spatial and scaling characteristics, making tumor segmentation in medical images a difficult task that might result in wrong diagnosis and therapy. Automation of a task like tumor segmentation is expected to enhance objectivity, repeatability and at the same time reducing turn around time. Conventional convolutional neural networks (CNNs) exhibit sub-par performance as a result of their inability to accurately represent the range of tumor sizes and forms. Developing on that, UNets have been a commonly used solution for semantic segmentation, and it uses a downsampling-upsampling approach to segment tumors. This paper proposes a novel architecture that integrates Attention-UNet with repeated Atrous Spatial Pyramid Pooling (ASPP). ASPP effectively captures multi-scale contextual information through parallel atrous convolutions with varying dilation rates. This allows for efficient expansion of the receptive field while maintaining fine details. The attention provides the necessary context by incorporating local characteristics with their corresponding global dependencies. This integration significantly enhances semantic segmentation performance. Our approach demonstrates significant improvements over UNet, Attention UNet and Attention UNet with Spatial Pyramid Pooling allowing to set a new benchmark for tumor segmentation tasks.
\end{abstract}

\section{Introduction}
\label{sec:intro}

An essential diagnostic tool for the evaluation, tracking, and planning of brain tumor surgery is magnetic resonance imaging (MRI). To highlight various tissue characteristics and tumor spread regions, multiple complementary 3D MRI modalities are typically acquired, that includes T1, T1 with contrast agent (T1c), T2, and Fluid Attenuation Inversion Recovery (FLAIR). In addition to saving doctors time, automated 3D brain tumor segmentation offers a precise, repeatable solution for additional tumor analysis and monitoring. Both in 2D natural images and in 3D medical image modalities, CNNs exhibit good segmentation accuracy and the ability to learn from examples.

The state-of-the-art architecture for
medical image segmentation is UNet \cite{10.1007/978-3-319-24574-4_28}, which combines
a symmetric encoder-decoder topology with skip-connections to
maximize information preservation. Li et al. \cite{li2018hdenseunet} developed a dense UNet for liver tumor segmentation from CT images. Higher segmentation accuracy was achieved by training neural networks with extraordinarily deep architectures, which was made possible by the emergence of residual CNNs \cite{inproceedings}. Combination of the techniques in \cite{Quan2016FusionNetAD} permit neural networks to have greater representational power facilitating capturing variations in complex data. The segmentation results can then be used to predict overall survival (OS) rate, instead where in earlier studies, machine learning models and radiomic features were combined to predict OS. Clustering plays a crucial role in MRI image processing, particularly for brain region segmentation and tissue analysis, as shown by Mirzaei and Adeli \cite{MirzaeiAdeli+2019+31+44}.

Due to pressing demands of treatment and similarity in appearance of healthy and affective tissues in an MRI image, CNN based methods are expected to capture more intricate details on intra-tumoral tissues.  The automatic segmentation of brain tumors has been made possible by the aforementioned CNNs, however, they still have the following two major issues: (i) CNNs often use repeated striding convolutional kernels and stacking numerous pooling layers to increase the receptive field for extracting additional features. Layer by layer, these processes lower the resolution of feature maps, which causes the tissue inside the tumor to lose small features and partial key as it propagates, (ii) It is crucial for CNN-based techniques to segment different tumor scales because brain tumors show varying sizes in MRI images. Nevertheless, the majority of CNNs now in use for brain tumor segmentation are only partially capable of multi-scale processing. Tumors can appear at different scales, locations, and shapes in medical images, making it difficult for UNet \cite{li2018hdenseunet} to detect and segment them accurately. As a result, these issues are now the primary barriers preventing the model from further enhancing brain tumor segmentation performance. To address these challenges, we propose a new framework motivated by \cite{10.6180--1830351969} and \cite{10.3389/fpubh.2023.1091850} involving ASPP which extracts features at multiple scales and locations, allowing the network to capture spatial heterogeneity and improve tumor segmentation accuracy. Recent advancements in medical image segmentation have seen the integration of Atrous Spatial Pyramid Pooling (ASPP) into UNet architectures, yielding improved results. Notably, Wang et al. \cite{WANG2021106268} proposed SAR-UNet, combining traditional U-Net with SE blocks and ASPP \cite{chen2017deeplab} to focus on relevant regions and extract multi-scale features, achieving significant results on the LITs dataset. Similarly, Ahmad et al. \cite{math10040627} adapted ASPP-UNet for Cardiac MRI segmentation. These studies demonstrate the potential of ASPP-enhanced U-Net architectures in accurately segmenting medical images, paving the way for further research and applications. Knowing UNet’s performance for semantic segmentation and the usefulness of ASPP, this paper proposes Attention based UNet with three ASPP blocks which has attention mechanism that helps the model to automatically learn the target structures of varying shapes and size.

\section{Method}
\label{sec:format}

\begin{figure*}[h!]
    \centering
    \includegraphics[height=9.5cm, width=11.5cm]{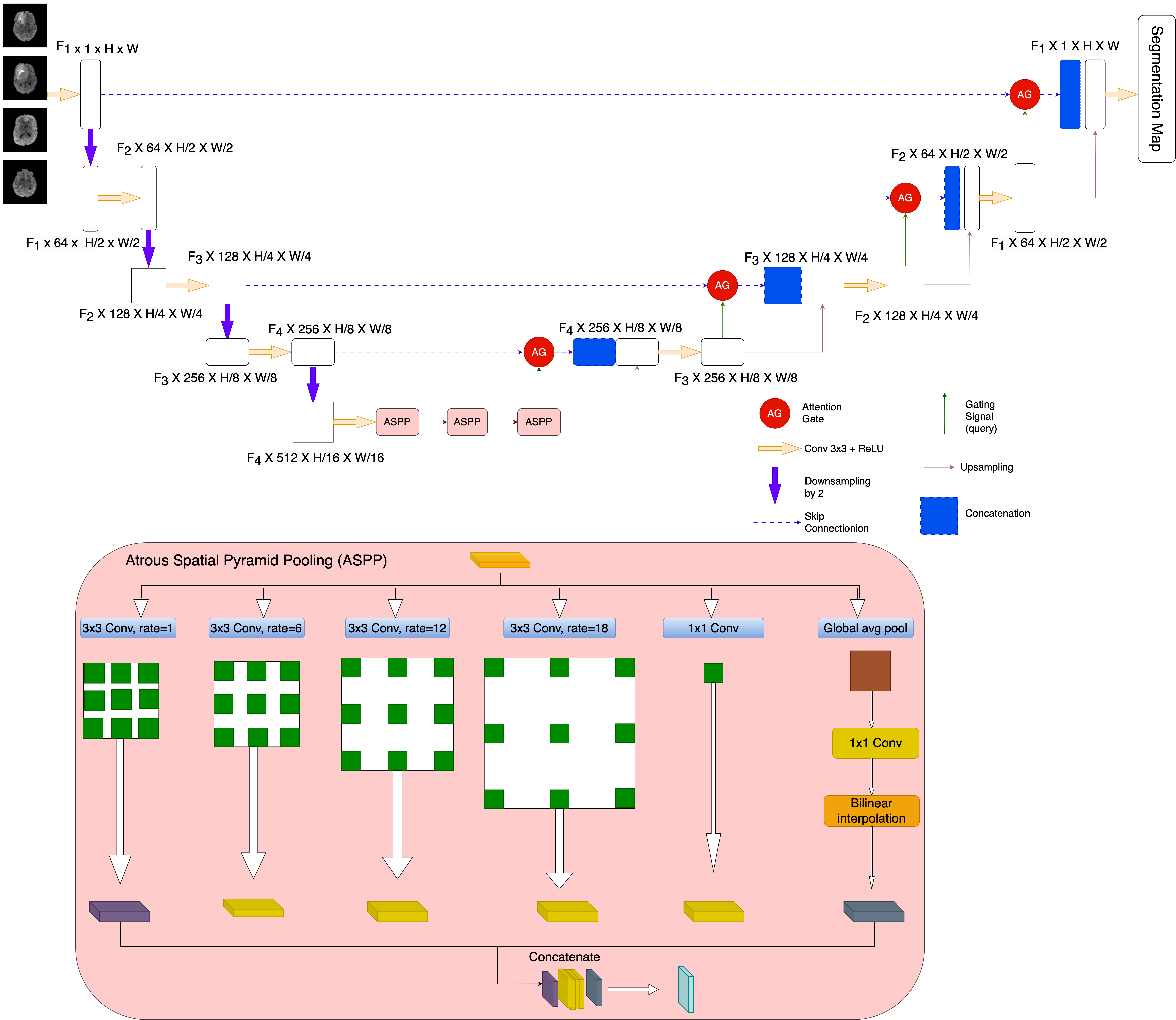}
    \caption{The end-to-end network architecture of the proposed framework. Input image is progressively filtered and downsampled by a factor of 2 at each level in the encoding part of the network. Attention gates (AGs) filter the features propagated through the skip connections. The Atrous Spatial Pyramid Pooling (ASPP) blocks further augment feature extraction with multiple convolution layers at varying dilation rates, capturing multi-scale contextual information.}
\end{figure*}

\begin{figure*}[h!]
    \centering
    \includegraphics[width=\textwidth]{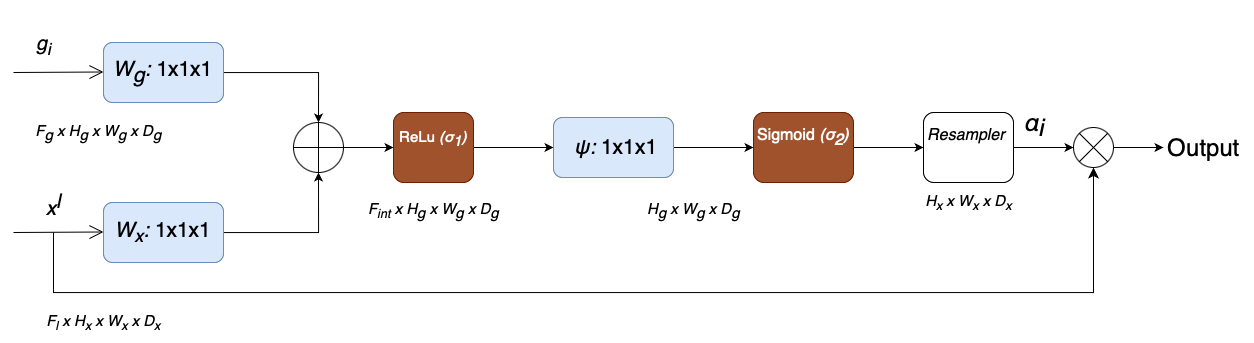}
    \caption{The schematic of attention gate (AG) \cite{article}. The $g_i$ and $x_l$ are represented as the gating signal vector and the feature map of the layer $l$, respectively. The $\sigma_1$ and $\sigma_2$ denote the Relu function and Sigmoid function. $W_x, W_g$ and $\psi$ are linear transformations. The $\alpha_i$ indicates the attention coefficient}
\end{figure*}
The proposed framework (Fig.1) integrates (i) four attention gates during upsampling and (ii) three ASPP blocks at the bottleneck of the network.
\subsection{BASE NETWORK}
\label{ssec:subhead}

The paper considers the basic UNet architecture and then build upon that by adding attention mechanisms as done by the authors in \cite{article}. ASPP blocks are integrated at the UNet bottleneck, capturing contextual information, increasing receptive field, and enabling long-range dependency analysis for enhanced segmentation accuracy.
\begin{table}
    \scriptsize 
    \centering
    \caption{Quantitative performance comparison of the proposed network (Att. UNet with ASPP) against the existing}
    \label{tab:my_label}
    
    \setlength{\tabcolsep}{0.4em} 
    \hspace*{-0.5cm} 
    \begin{tabular}{|l|ccc|ccc|ccc|}
    \hline
        & \multicolumn{3}{c|}{T1C} & \multicolumn{3}{c|}{T2 FLAIR} & \multicolumn{3}{c|}{T2W}  \\
        \cline{2-4} \cline{5-7} \cline{8-10}
        Model & DSC & mIoU & Acc & DSC & mIoU & Acc & DSC & mIoU & Acc \\
        \hline
        UNet & 63.15 & 36.74 & 98.58 & 32.44 & 16.22 & 94.60 & 70.33 & 38.59 & 99.02 \\
        Att. UNet & 64.71 & 37.91 & 98.64 & 78.30 & 44.31 & 99.15 & 71.63 & 38.73 & 98.91\\
        Att. UNet with SPP & 65.87 & 38.89 & 98.72 & 75.39 & 41.27 & 99.13 & 72.57 & 41.45 & 98.92\\
        Att. UNet with ASPP & \textbf{68.76} & \textbf{41.13} & \textbf{98.74} & \textbf{79.75} & \textbf{45.83} & \textbf{99.17} & \textbf{75.46} & \textbf{42.49} & \textbf{99.04} \\
    \hline
    \end{tabular}
\end{table}

\begin{table*}[ht!]
    \centering
    \scriptsize 
    \caption{Table showing the model along with their respective training and inference time.}
    \label{tab:performance_comparison}
    \setlength{\tabcolsep}{0.6em} 
    \renewcommand{\arraystretch}{1.2} 
    
    \begin{tabular}{|l|cc|cc|cc|}
    \hline
        & \multicolumn{2}{c|}{\textbf{T1C}} & \multicolumn{2}{c|}{\textbf{T2 FLAIR}} & \multicolumn{2}{c|}{\textbf{T2W}} \\
        \cline{2-3} \cline{4-5} \cline{6-7}
        \textbf{Model} & \textbf{Training Run Time} & \textbf{Inference Time on Test Set} & \textbf{Training Run Time} & \textbf{Inference Time on Test Set} & \textbf{Training Run Time} & \textbf{Inference Time on Test Set} \\
    \hline
        UNet & 141min 15.6s & 13.6s & 141min 27.5s & 13.8s & 26min & 37.1s \\
        Att. UNet & 135min 35.5s & 14.5s & 131min 25s & 14.3 & 29min 38.5s & 13s \\
        Att. UNet with SPP & 129min 53.4 & 31.6 & 79min 11.2s & 31.4s & 113min 46.5s & 30.7s \\
        Att. UNet with ASPP & 45min 35.1s & 13.2s & 132min 36.9s & 14.7s & 31min 17.2s & 13.2s \\
    \hline
    \end{tabular}
\end{table*}
\subsection{ATTENTION GATE}
\label{ssec:subhead}
In conventional CNN designs, the feature-map grid is gradually downsampled in order to capture a broad enough receptive field and, consequently, semantic contextual information. Features at the coarse spatial grid level thereby model the global tissue location and connection. Reduction in false-positive predictions for small objects with high shape diversity remains a challenging task. Current segmentation frameworks \cite{khened2019fully,Roth2017Hierarchical3F,Roth2018} reduce the task into distinct localization followed by segmentation steps in order to increase accuracy by relying on prior object localization models. Here, we show that including attention gates (AGs) into a UNet can accomplish the same goal. It is not necessary to train several models or a significant amount of additional model parameters for this. Unlike multi-stage CNNs' localization paradigm, AGs gradually reduce feature responses in background regions that are irrelevant, eliminating the need to crop a ROI between networks.

From Fig.2, each attention gate uses a gating vector $g_i \in \mathbb{R}^{F_{g}}$ for each pixel $i$ in the feature map that contains contextual information to determine which regions of input the feature maps should be focused on. This gating vector helps in pruning lower-level feature responses that are not relevant to the task at hand.
The attention coefficients $\alpha_i \in [0,1]$ are computed using an additive attention mechanism, which is formulated to produce a scalar value for each pixel. This scalar value indicates the importance of that pixel in the context of the task. Additive attention is used to obtain the gating coefficient and is formulated as   \cite{Bahdanau2014NeuralMT}:
\begin{align}
&q_{\mathit{att}}^l=\psi^T(\sigma_1(W_x^Tx_i^l+W_g^Tg_i+b_g))+b_{\psi} \label{eq1}\\
&\alpha_i^l=\sigma_2(q_{\mathit{att}}^l(x_i^l, g_i; \Theta_{\mathit{att}}) \label{eq2}
\end{align}
Where $\sigma_1$ is often chosen as ReLu function, $\sigma_1(r)=max(0,r)$, $\sigma_2(r)=\frac{1}{1+e^{-r}}$ correspond to sigmoid activation function and $q_{\mathit{att}}^l$ is the attention query. The final output of the attention gate is obtained by element-wise multiplication of the input feature maps and the attention coefficients ${\hat{x}^l}_{i,c}=x^l_{i,c} \cdot \alpha^l_i$. This process allows the model to selectively enhance features that are deemed important while suppressing those that are not. The result of multiplication will reduce the value of the
unrelated region and increase the value of the target region,
thus improving the segmentation accuracy.
AG's are characterized by a set of parameters $\Theta_{\mathit{att}}$ containing linear transformations $W_x \in \mathbb{R}^{F_l \times F_{int}}$, $W_g \in \mathbb{R}^{F_g \times F_{int}}$, $\psi\in \mathbb{R}^{{F_{int}} \times 1}$ and bias term $b_\psi \in \mathbb{R}$, $b_g \in \mathbb{R}^{F_{int}}$. The linear transformations are computed using
channel-wise 1x1x1 convolutions for the input tensors.

\subsection{ATROUS SPATIAL PYRAMID POOLING}
\label{ssec:subhead}
In the semantic segmentation task using ASPP as described in \cite{chen2017deeplab}, it is possible to generate larger
receptive field for the features extracted from the images, and hope that the resolution of feature images
will not decrease too much (too much resolution loss will lose a lot of detailed information about the
image boundary), but these two are contradictory. If larger receptive fields are to be obtained, larger
convolution kernels or larger steps are needed for pooling. For the former, the amount of calculation is
too large, and the latter will lose resolution. The dilated convolution is used to solve this
contradiction, and it can get a larger receptive field without losing too much resolution. The dilated
convolution is shown in Fig.1 ASPP Block. Compared with ordinary convolution, atrous convolution enlarges the
receptive field when the pooling loss information is small, so that each convolution output contains a
wider range of information, ensuring that the output image features have a higher resolution. The equation for atrous convolution is:
\begin{align}
    &y[i]=\sum_{k=1}^{K} x[i+r \cdot k]w[k]\label{eq3}
\end{align}
When $r=1$ it is the standard convolution and when $r>1$ it is the atrous convolution which is the stride to perform the sampling in the input space during convolution.
Atrous convolution allows an efficient mechanism to control the field-of-view and finds the best trade-off between accurate localization (small field-of-view) and context assimilation (large field-of-view).
In ASPP, parallel atrous convolution with different dilation rate are applied to the input feature map and are then fused together. As objects of the same class can have different scales in the image, ASPP helps to account for different object scales which improves the accuracy of segmentation. 
\begin{figure*}[h!]
    \centering
    \includegraphics[width=1\linewidth]{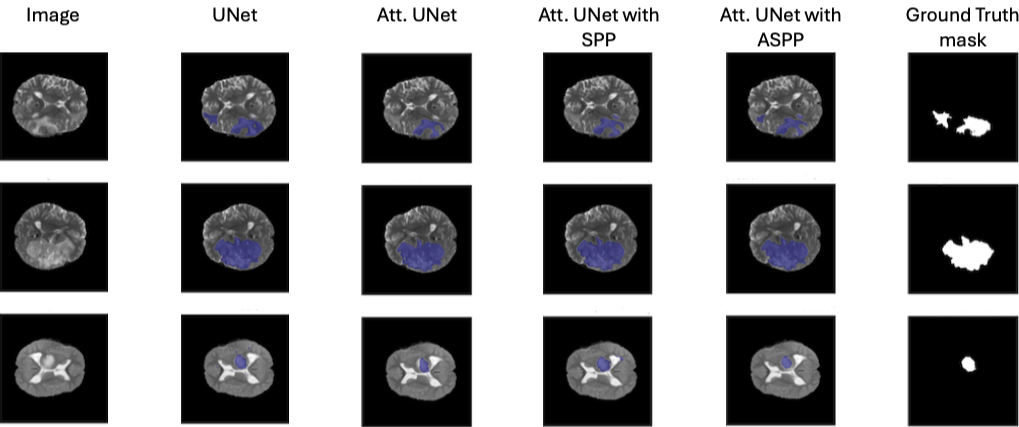}
    \caption{ Visual comparison of segmentation output from UNet, Attention UNet, Attention UNet with SPP and Attention UNet with ASPP on T2W images.}
\end{figure*}

\section{EXPERIMENTS AND RESULTS}
\label{sec:pagestyle}
\subsection{DATASET}
\label{ssec:subhead}
The BraTS 2023 Adult Glioma dataset \cite{kazerooni2024braintumorsegmentationbrats} is used in this study. This dataset has MRI images from 1,251 patients. The images have dimensions 240 × 240 × 155 where 240 denotes the height and width and 155 is the number of slices or the number channels in the z direction since its a volumetric data and comprises of the voxel. Nineteen institutions provided the data, which are
gathered using a variety of MRI scanners. Upon analyzing the dataset, it is observed that the axial planes offer enhanced visibility of tumor boundaries and diffusion to other regions of the brain, which is expected to contribute to more accurate segmentation of the tumor. Therefore, we extract the axial slices of each T1 contrast, T2 weighted and FLAIR (Fluid Attenuated Inversion Recovery) from the data of every patient. 
Three datasets are created for model training and evaluation, comprising 1000 training, 125 validation and 126 test images. Training and validation sets include $(image, mask)$ pairs, while test images are used for segmentation mask prediction and evaluation.

\subsection{IMPLEMENTATION DETAILS}
\label{ssec:subhead}
All the images are made sure to have the same dimension of $240\times240$. This architecture of the network is implemented on the Pytorch platform with NVIDIA GeForce RTX 3090 GPU. The  batch size is set as 32 for both training and validation, and the ADAM optimizer is used with an initial learning rate of $1\times10^{-3}$ for 100 epochs. 
Cosine Annealing Scheduler is used as a learning rate scheduling technique that adjusts the learning rate during training using a cosine annealing schedule. Cosine annealing permits a first starting rate with a high learning rate that is gradually decreased to a minimum value before gradually increasing again. Mathematically it is:
\begin{align}
    &\eta_t = \eta^{i}_{min} + \frac{1}{2}(\eta^{i}_{max}-\eta^{i}_{min})(1+cos(\frac{T_{cur}}{T_i})\pi)\label{eq4}
\end{align}

Where $\eta^{i}_{min}$ and $\eta^{i}_{max}$ are ranges for the learning rate, $T_{cur}$ account for how many epochs have been performed since the last restart and $T_{i}$ is the total number of epochs in the current cycle.

Cosine annealing mimics the cooling process in simulated annealing \cite{doi:10.1126/science.220.4598.671} by gradually reducing the learning rate following a smooth cosine curve. This allows for effective exploration at higher rates in the beginning and fine-tuning at lower rates later, enabling the model to converge to better minima. Its gradual and cyclical adjustments prevent abrupt changes, making it particularly effective for complex tasks like brain tumor segmentation, where precise optimization is crucial.

\subsection{COMPARISON AND RESULTS}
\label{ssec:subhead}
The results demonstrate the effectiveness of our proposed framework, which combines attention mechanisms and Atrous Spatial Pyramid Pooling (ASPP) to improve brain tumor segmentation. By leveraging the synergies between these two techniques, our framework effectively captures spatial heterogeneity and contextual information. The experimental results presented here showcase the framework's ability to precisely segment brain tumors with diverse characteristics, highlighting its potential for clinical applications.

T1C images improve tumor visibility, but introduce variability due to the timing of contrast agent administration and patient-specific factors. This variability, coupled with noise and artifacts common in T1C images, can obscure tumor boundaries and mislead the model. Furthermore, despite attention mechanisms and ASPP blocks that improve feature extraction, the model can struggle if T1C specific features are underrepresented in the training data, leading to suboptimal performance of the model on T1C images.

\section{Conclusion}
\label{sec:typestyle}

In conclusion, the proposed framework makes a significant improvement in brain tumor segmentation, leveraging the synergies between attention mechanisms and ASPP to effectively capture spatial heterogeneity and contextual information. This integration enables accurate segmentation of brain tumors with diverse characteristics, as evidenced by experimental results presented here. Notably, the proposed model consistently outperforms UNet, Attention UNet and Attention UNet with Spatial Pyramid Pooling as shown in Table 1 across key evaluation metrics such as DSC and IoU.
These advancements have profound implications for clinical practice, enhancing segmentation accuracy and potentially leading to more reliable diagnoses and treatment plans. Future research directions will focus on refining the model and exploring its applicability to other medical imaging tasks, with the ultimate goal of improving diagnostic efficacy of the patient. 


\bibliographystyle{IEEEbib}
\bibliography{strings,refs}

\end{document}